\documentclass[12pt,showpacs,amssymb]{iopart}
\usepackage{dcolumn}  
\usepackage{bm}       
\usepackage[dvips]{graphicx}

\newcommand{\beq}{\begin{equation}}
\newcommand{\eeq}{\end{equation}}
\newcommand{\Matlab}{\textsc{Matlab}}
\newcommand{\Comsol}{\textsc{Comsol}}

\newcommand{\nbf}{\mathbf{n}}

\newcommand{\vbf}{\mathbf{v}}
\newcommand{\ubf}{\mathbf{u}}
\newcommand{\tbf}{\mathbf{t}}
\newcommand{\zerovec}{\mathbf{0}}

\newcommand{\nablabf}{\boldsymbol{\nabla}}

\newcommand{\gammabf}{\boldsymbol{\gamma}}

\newcommand{\Ubf}{\mathbf{U}}

\newcommand{\intd}{\mathrm{d}}

\newcommand{\Nb}{N_{\beta}}

\newcommand{\refp}{p_{\mathrm{ref}}}

\begin{document}

\title[Optimally homogenized Bio-reactors]
{Optimal homogenization of perfusion flows in microfluidic bio-reactors; a numerical study}

\author{Fridolin Okkels, Martin Dufva, and Henrik Bruus}

\address{Department of Micro- and Nanotechnology, Technical University of Denmark, \\
DTU Nanotech, Building 345 East, DK-2800 Kongens Lyngby, Denmark}

\ead{fridolin.okkels@nanotech.dtu.dk}

\date{March 24$^\mathrm{th}$ 2009, ver. 14}

\begin{abstract}
To ensure homogeneous conditions within the complete area of
perfused microfluidic bio-reactors, we develop a general design of a
continuously feed bio-reactor with uniform perfusion flow. This is
achieved by introducing a specific type of perfusion inlet to the
reaction area. The geometry of these inlets are found using the
methods of topology optimization and shape optimization. The results
are compared with two different analytic models, from which a
general parametric description of the design is obtained and tested
numerically. Such a parametric description will generally be
beneficial for the design of a broad range of microfluidic bioreactors used for e.g.~cell culturing and analysis, and in feeding bio-arrays.
\end{abstract}

\submitto{\JMM}
\pacs{87.85.M-, 02.60.Pn, 47.15.-x, 87.85.Va}
\maketitle

\section{Introduction}
 \label{sec:introduction}
The development of microfluidics, to handle minute amounts of
fluids, is currently revolutionizing fluid transport in the field of
analytic cell-biology: Traditionally, cells are cultured in
so-called batch cultures in a flask and an experiment is typically
initiated by adding an agent. After a certain time, such as a day or
two, the response of the agent is studied using typically only one
reporter such as fluorescence. In order to increase throughput,
cells can, at present, be cultured and assayed in robotically
controlled 96 or 384 well plates. By contrast, culturing of cells on
a microfluidics device gives a range of new possibilities
\cite{ElAli:06a} e.g.~studying cell mobility in real time when
exposed to stable continuous gradients \cite{LiJeon:02a}.
Furthermore, combinatory experiments can be performed on chip that
are based on arrays of interconnecting chambers
\cite{Hung:05a,Hung:05b}.

The inlet design presented in this paper introduce a number of
improvements to current perfused bio-reactors: The creation of
uniform flow conditions all over the bio-reactor ensures homogeneous
cell conditions both with regards to concentrations of externally
supplied growth-factors and to the shear induced on the cells by the
perfusion flow. Too small a height of cell culture chips is inhibiting cell growth \cite{Yu:05a,Petronis:06a,Korin:07a}, and in Refs.~\cite{Stangegaard:06a,Stangegaard:06b} it has been shown that the chamber height must exceed 1.5 mm in order to provide identical culturing conditions as in traditional cell culture flask. On the other hand, to ensure laminar flow conditions, a small height
is preferred. Therefore in the case of cell-culturing chips, a
chamber height of 1.5 mm is optimal. In other cases, such as for
micro-array hybridization chambers, the functionality indeed benefit
from far smaller reactor-channel heights, where volume needs to be
minimized in combination with a maximization of reaction area.

In recent years different bio-reactors have been constructed, where
the uniformity of the perfusion flow along the reaction area has
been achieved at the expense of a large hydraulic resistance across
the whole bio-reactor \cite{Hung:05a}. One example is the Micro
cell-culture chamber by M.~Stangegaard et al.~\cite{Stangegaard:06a}, where the fluid is directed from a wide
reservoir through a large number of small parallel channels. This
barrier creates a large pressure drop which give rise to the uniform
flow. From the inlet structure described from this work, the same
uniform flow-field can be achieved with a significantly lower
pressure-drop. This opens up the possibilities of driving the
perfusion flow by low-power methods such as e.g.~buoyancy force,
which recently has been used to drive other microfluidic devices
\cite{NatConvect}.

Our novel design also reduces the fluid volume used in creating the
uniform flow, which is crucial both when dealing with expensive
biochemical samples or to avoid dilution of small samples.
Additionally it will enable a better analysis of fast cell-reaction
kinetics with high time resolution.

The paper is organized as follows: In section~\ref{sec:Lauout} the
general bio-reactor layout is outlined together with an introduction
of the related characteristic parameters. In
section~\ref{sec:Optimization} the optimal structure of the
perfusion inlet is found, first by the general method of
topology optimization, which imposes no constraints on the topology of the structure. The resulting structure is further refined by the
method of shape optimization. The optimized geometry is in
section~\ref{sec:AltGeom} compared to two simple expansion design,
while in section~\ref{sec:Guide} the results are summarized in a
design guide. The analysis and the design guide is further verified by full 3D simulations in section~\ref{sec:Sim3D}. A conclusion is given in sec.~\ref{sec:Conclusion}.

\begin{figure}
\centering
\includegraphics[width=10cm,clip]{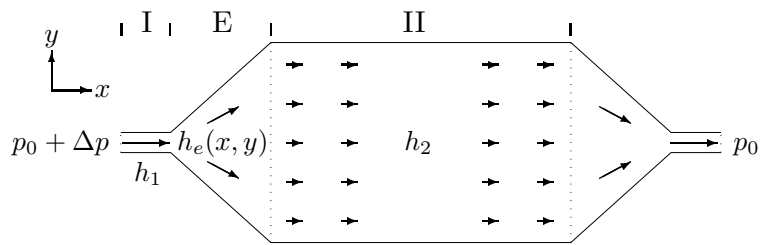}
\caption{A sketch of the bio-reactor with its three sections: (I)
the perfusion inlet of constant height $h_1$, (E) the expansion
chamber of spatially varying height $h_e(x,y)$, and (II) the main
reactor area of constant height $h_2$. The perfusion flow, driven by
the pressure drop $\Delta p$, is indicated by arrows.}
\label{fig:Overview}
\end{figure}

\section{Geometry and basic flow equations of the microfluidic bio-reactor}
 \label{sec:Lauout}
The generic microfluidic bio-reactor layout used in this work is
illustrated in figure~\ref{fig:Overview}. It consists of a single
microchannel perfusion inlet (I) of constant height $h_1$, which
broaden out in an expansion chamber (E) of varying height $h_e(x,y)$
to distribute the fluid over the much wider and more shallow main
reactor (II) of constant height $h_2$, where the cells are
immobilized. All vertical channel and chamberheights in the $z$
direction are much smaller than any lateral length scale in the $xy$
plane; the bio-reactor is thus flat.

The main objective is to obtain a uniform flow in the main reactor
with minimal pressure drop $\Delta p$ and with a minimal volume of
the expansion chamber. This is achieved by carefully designing
variations in chamber height $h_e(x,y)$ of the expansion chamber. As
the constant inlet channel height $h_1$ is assumed larger than the
constant height of the main reactor area $h_2$, the height variation
in the expansion chamber (E) will be bounded by these two heights:
 \beq
 \label{eq:he_limit}
 h_1 \ge h_e(x,y) \ge h_2.
 \eeq
The whole bio-reactor is assumed symmetric both through a central
vertical and a central horizontal axis, and as a consequence only
the upper left part will be dealt with here.

As we consider only low concentrations of the solutes and a constant
temperature, the density $\rho$ and viscosity $\eta$ of the
buffer liquid are constant in space, and the flow is determined by
the geometry of the reactor and the applied pressure drop $\Delta p$
driving the flow. As a consequence of the assumed flatness of the
bio-reactor, the pressure $p$ does not vary in the vertical $z$
direction, i.e.\ $\partial p/\partial z = 0$. Moreover, due to the
small heights, viscous damping from the top and bottom plates of the
bio-reactor dominates the fluid flow and makes the flow laminar.
This is evident from the value of the Reynolds number $Re$
given the low flow velocities, $u \approx 1$~mm/s, and small
length-scales, $h \approx 1$~mm of the system: $Re \approx 1$.

In this flow regime it is useful to work with the $z$-averaged 2D
velocity field $\ubf(x,y) = \big(1/h(x,y)\big)\int_0^h
\vbf(x,y,z)\:\mathrm{d}z$ of the full 3D field $\vbf(x,y,z)$.
To a good approximation $\ubf$ fulfills the 2D Brinkman-Darcy equation
\cite{BruusBook},
 \beq
 \label{eq:Brinkman_eq}
 \eta \nabla^2\ubf - \frac{12\eta}{h^2(x,y)}\,\ubf - \nablabf p(x,y)
 = \zerovec.
 \eeq
Here the prefactor $12\eta/h^2(x,y)$, also denoted the damping
coefficient $\alpha$,
 \beq
 \label{eq:Lubric_rel}
 \alpha = \frac{12\eta}{h^2(x,y)},
 \eeq
is reminiscent of the $z$-part of the Laplace operator in the full
3D-description, and it represents the dominant part of the viscous
damping of the liquid in the system.

For possible continues changes in the
height $h(x,y)$ of the expansion region (E), the $z$-averaged 2D
velocity field is not divergence-free due to mass-conservation, but
an additional tern arises: $\nablabf\cdot\ubf(x,y) = -|\nablabf
h(x,y)|/h(x,y)$. In the case of possible discontinues jumps in
height along an interface, this correction becomes the following new
boundary condition on the interface:
 \beq
 \label{eq:BC_HeightJump}
 h_1\ubf_1\cdot\nbf = h_2\ubf_2\cdot\nbf, \quad
 \ubf_1\cdot\tbf = \ubf_2\cdot\tbf, \quad p_1 = p_2,
 \eeq
where the height-subscript is extended to the corresponding
velocities and pressures.

\begin{figure} 
\centering
\hspace*{21mm}
\includegraphics[height=45mm,clip]{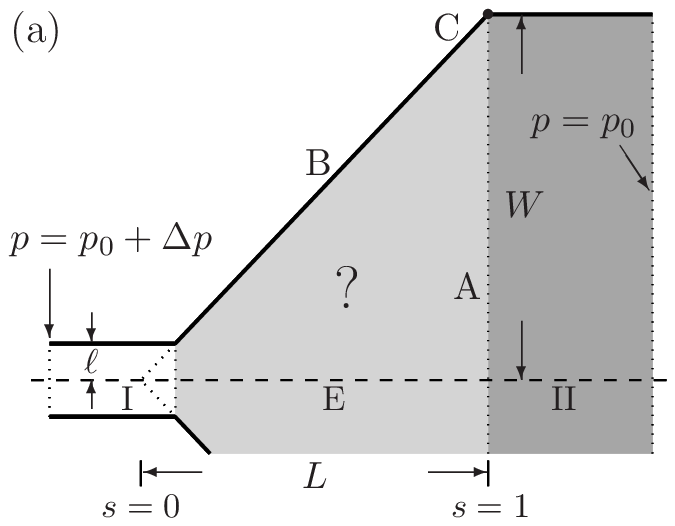}
\hspace*{5mm}
\includegraphics[height=45mm]{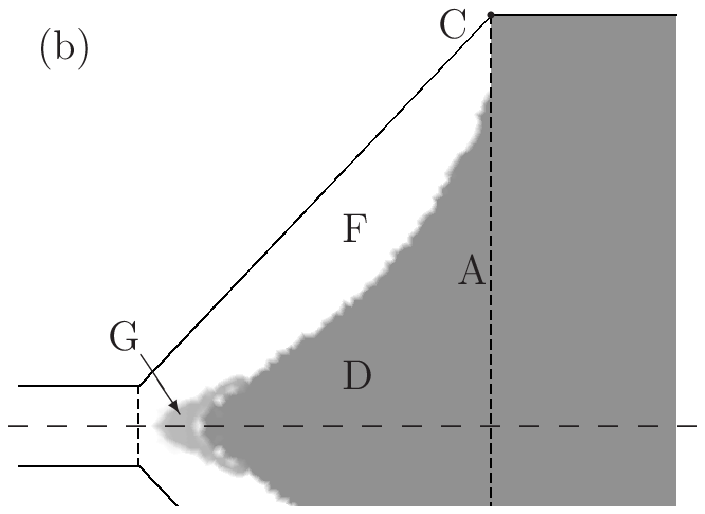}
\caption{(a) The geometry of the expansion chamber E. The parameters
related to geometry and flow are defined together with specific
segments and points used for later reference. The optimal transition
between the inlet height $h_1$ (white area) and the main reactor
channel height $h_2$ (gray) occurs inside the expansion region E
(light gray). (b) Topology optimized height variation in the
expansion region. The height changes almost step-like from the white
region F ($h_1$) to the gray region D ($h_2$). Note the small region
of intermediate height marked G.}
 \label{fig:Setup_G}
\end{figure}

Working with the 2D-restricted description, the detailed geometry of
the expansion chamber is illustrated in figure~\ref{fig:Setup_G}.
The important in-plane length scales are the length $L$ of the
expansion chamber, the width $W$ of the main reactor, and the width
$\ell$ of the inlet. To achieve a uniform flow in the main chamber,
the pressure along the line A dividing the expansion chamber and the
main chamber must by constant. Consequently, the spatial variation
of the expansion chamber height $h_e(x,y)$ must be optimized in
order to get as homogeneous pressure along line A as possible.

%
%

%
\section{Optimization}
 \label{sec:Optimization}
To enable optimization of the system two additional parts are
introduced. First, a set of design variables $\gammabf$, which
uniquely characterizes all available configurations in the
optimization problem, and for which a unique solution
$\Ubf(\gammabf)$ to the system exists. Second, an objective function
$\Phi$ which quantifies how well a given configuration of the system
performs. By convention $\Phi$ has to be minimized in order to
achieve the optimal solution, and generally the objective function
can depend on the design variables and the related solution of the
system $\Phi = \Phi\Big(\gammabf,\Ubf(\gammabf)\Big)$.

As alluded to in the previous section, we base our objective
function on the homogeneity of the pressure along the cross-section
A, since a uniform pressure there will lead to the required uniform
flow field in the main reactor. In the following, this objective
will be expressed in two different ways, depending on the given
optimization methods.
\subsection{Topology optimization of the spatial height variation}
To search for the globally optimal solution, and not a priori
exclude any non-intuitive solutions, we will not rely on any
pre-described variation of the height. Therefore, we begin by
applying the method topology optimization \cite{Bendsoe03}, which by
definition is independent of the topology and therefore unlimited in
its search for the optimal bio-reactor design. The method of
topology optimization was first applied to the field of structural
mechanics\cite{Bendsoe88}, and have been recently implemented to the
field of microfluidic systems \cite{Borrvall,Olesen:06a} and
chemical microreactors \cite{Okkels:07a}.

Arbitrary height variations of $h_e(x,y)$ can be realized by
representing the height as a variation of the design variable field
$\gamma(x,y)$, where $0 \leq \gamma \leq 1$. To cover the range of
heights defined in equation~(\ref{eq:he_limit}), the design variable
is assigned the value $\gamma = 0$ to describe chamber heights equal
to the inlet height $h_1$ and the value $\gamma = 1$ for heights
equal to the main reactor height $h_2$. In the expansion chamber,
now denoted the design region $\Omega$, the design field can take
any value $0\leq \gamma(x,y) \leq 1$ to describe all possible height
variations $h_e(x,y)$ .

The actual implementation, method and procedure of topology
optimization will not be touched upon here, as it is fully described
in the work of Olesen, Okkels and Bruus \cite{Olesen:06a}. Still
what is essential for this work is the objective function $\Phi$,
which has to be chosen with care. To obtain a numerically stable
search we define $\Phi$ as the square deviation of the pressure
around a reference pressure $\refp$ along A:
 \beq
 \label{eq:ObjFunc_TOPOPT}
 \Phi = \frac{1}{W}\int_A(p-\refp)^2 \intd s.
 \eeq
We choose $\refp$ as the pressure at the far corner C in the case
where the expansion chamber has the same height $h_1$ everywhere as
the inlet.

Figure~\ref{fig:Setup_G}(b) shows the resulting optimal height
distribution $h_e(x,y)$ for the following set of parameters:
$W=10^{-2}\,\mathrm{m}$, $L=0.95\,W$, $\ell = 0.1\,W$, $h_1 =
0.04\,W$, and $h_2 = 0.02\,W$, where the gray-scale color-coding
spans from height $h_1$ in white down to $h_2$ in gray. We define
the ratio $A$ between the two damping coefficients $12\eta/h_1^2$
and $12\eta/h_2^2$ as
 \beq
 \label{eq:Def_A}
 \mathcal{A} = \left(\frac{h_1}{h_2}\right)^2,
 \eeq
and get the value of $\mathcal{A}=4$ for figure
\ref{fig:Setup_G}(b).

As mentioned earlier, any change in height
produces a correction-term to the continuity equation, when using a
2D-restricted description. It turns out that both ways of
implementing this correction in the topology optimization problem
fails, due to the very nature of the method. First, the free
variations of the design field in topology optimization prohibits
any interface to be defined a priori, and therefore the boundary
conditions of equation (\ref{eq:BC_HeightJump}) cannot be applied in
this step of the optimization procedure. Second, it turns out that
the solutions of the topology optimization problem involves sharp
transitions in the height, limited by the grid-meshing length-scale
of the finite element method. Therefore when including the
correction-tern to the continuity equation, a fluid source is added
to single mesh-elements, and this destabilizes the convergence of
the method. The way to work about this limitation, is to add the
boundary conditions of equation (\ref{eq:BC_HeightJump}) to the
shape-optimization method, applied later in the optimization
process, after the shape-optimization has been preliminarily
compared to the topology-optimization.
%

From the topology optimized solution in figure~\ref{fig:Setup_G}(b)
we see that among all possible height variations, the optimal design
consists of a single sharp transition between a region of inlet
height $h_1$, and a region of main reactor height $h_1$. Only very
close to the inlet channel is seen an ambiguity which indicate the
possible existence of a region of intermediate height. From topology
optimizations for other parameter-values, similar solutions arise
with a sharp transition between regions of height $h_1$ and $h_2$,
and consequently we conclude that such a single-connected transition
is indeed the optimal solutions of the problem.

\begin{figure} 
\centering
\includegraphics[height=5cm]{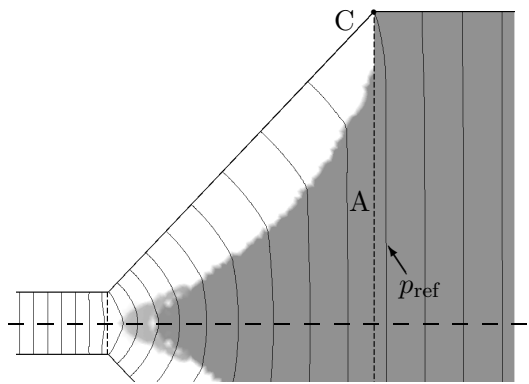}
\caption{The contour lines of the pressure $p(x,y)$ in the
topology optimized design. The contour $p=\refp$, going through
the corner C is marked.} \label{fig:TOPOPT_P}
\end{figure}

To evaluate the quality of the topology optimized solution, we plot
in figure~\ref{fig:TOPOPT_P} the pressure contour lines of the
solution in figure~\ref{fig:Setup_G}(b), including a contour line
corresponding to the value $p = \refp$ which goes through the
corner $C$ of the expansion chamber. Except close to the upper
side wall, the pressure is seen to be uniform at the entrance of the
main reactor, and it decreases uniformly throughout the whole
extension of the main reactor.
%

The chosen parameters used for the solution in
figures~\ref{fig:Setup_G}(b) and~\ref{fig:TOPOPT_P} represent a
rather extreme case, i.e.,~a combination of a small
height-difference $\mathcal{A}= 4$ and a wide expansion $L/W =
0.95$. As a result the transition extends nearly through the whole
expansion region, but for all common parameters, the type of
solution remains optimal.

From the results of the topology optimization it is therefore
natural to proceed with the shape optimization method, which
compared to topology optimization involves fewer design parameters,
is faster, and is numerically more stable.

\subsection{Shape-optimization}
\begin{figure} 
\centering
\hspace*{24mm}
\includegraphics[height=43mm,clip]{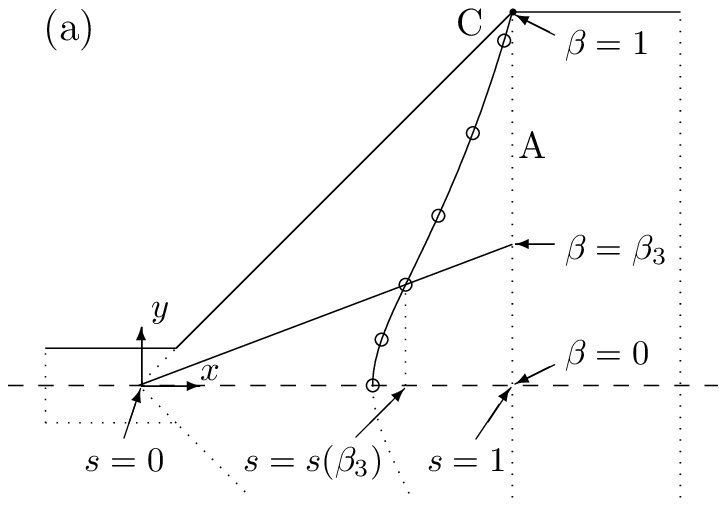}
\includegraphics[height=43mm,clip]{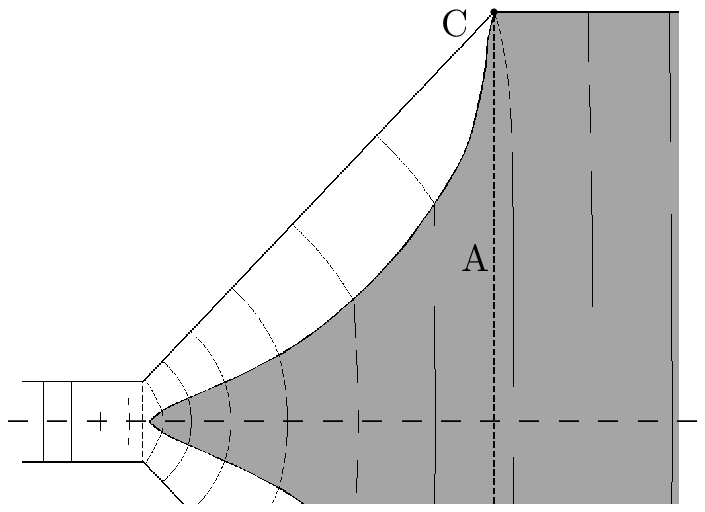}
\caption{(a) Setup for the parametrization of the shape-optimization
problem. Variables $s$ and $\beta$ parameterize the normalized $x$
and $y$-axis respectively, and the control-points $s(\beta_i)$ are
shown by circles with the interpolated interface curve in solid.
(b) The pressure contours (thin lines) of the shape optimized
positioning of the interface line $(s,s\beta)$ (thick line).
Similarly to figure~\ref{fig:TOPOPT_P} the pressure contour are
originating from the upper corner C} \label{fig:Setup_fb}
\end{figure}

In shape optimization the interface line between the heights $h_1$
and $h_2$ in the transition chamber is given by a cubic
interpolation line through a number of control points $(x_i,y_i),
i=1,2,\ldots,\Nb$ as shown in figure~\ref{fig:Setup_fb}(a) with
$\Nb=6$. It is convenient to parameterize the points by the
expression
 \beq
 \label{eq:Expand_Param}
 (x_i,y_i) = (s_i L,\,s_i \beta_i W),
 \qquad \frac{\ell}{W}<s_i<1, \quad 0<\beta_i<1,
 \eeq
which ensures that all points lie within the expansion chamber.
By fixing the factors $\beta_i$ by $\beta_i =
[1-1/(3\,\Nb)]\times[(i-1)/(\Nb-1)]$, a relatively even distribution
of the control points is also ensured. During the optimization
process the position of the interface line is changed by adjusting
the control points $s_i$.

The optimization is carried out by a simplex-method relying only on values
of the objective function $\Phi(\gammabf)$, and not its partial
derivatives $\partial \Phi/\partial \gammabf$. To ensure efficient
convergence of the given simplex method, it is beneficial to assign
initial values around unity for the design variables $\gammabf$.
Furthermore, the method is unbound i.e.~the design variables must
give rise to a well-defined geometry regardless of their value. All
this is accomplished by using the arcus tangent function:
 \beq
 \label{eq:DefOptParam}
 s(\beta_i) = 1 - \gamma_0 \,(1-\beta_i)
    \left\{ 1+\frac{2}{\pi}\arctan\left[\frac{\pi}{2}(\gamma_i-1)\right] \right\},
    \qquad i=1\ldots \Nb,
 \eeq
with
 \beq
 \gammabf = \{\gamma_0,\gamma_1,\ldots,\gamma_{\Nb}\}.
 \eeq
We use $\Nb+1$ design variables to determine $\Nb$ shape parameters
because a faster convergence is achieved by adjusting the extend of
the whole interface by a single parameter $\gamma_0$. Furthermore,
equation~(\ref{eq:DefOptParam}) let the initial configuration of
 \beq
 \label{eq:Init_Sh_gamma}
 \gammabf_\mathrm{init} = [\gamma_0,1,\ldots,1]
 \eeq
give rise to a well-defined, straight interface line reaching from
the position $(x,y)=\Big((1-\gamma_0)L,0\Big)$ to the upper corner.

Now that the interface by definition extends to the upper corner,
this constraint does not need to be included in the objective
function $\Phi$ of the shape-optimization, and $\Phi$ can therefore be defined
with the sole purpose of achieving a uniform pressure along segment A:
 \beq
 \label{eq:ObjFunc_ShOpt}
 \Phi = \frac{1}{W}\int_A
   \left|\frac{\partial p}{\partial y}\right| \intd s.
 \eeq

The actual optimization of the design-variables follows two steps:
First a rough initial interface is found by using the initial setup
of equation~(\ref{eq:Init_Sh_gamma}), and only adjusting the single
variable $\gamma_0$, using a simple \Matlab{} implementation of a
bounded golden section search with combined parabolic interpolation
\cite{fminbnd_Search}. Once a suitable straight initial interface
is found, the actual shape is obtained using a direct unbounded
simplex search method, also implemented in \Matlab{}
\cite{fminsearch_Search}.

\subsection{Results}
First, the shape-optimization method has to be validated with respect
to the topology optimized solution, shown in
figures~\ref{fig:Setup_G}(b) and~\ref{fig:TOPOPT_P}. The same
parameter-values were used, and the resulting shape-optimization
shown in figure~\ref{fig:Setup_fb}(b), is indeed similar to
figure~\ref{fig:TOPOPT_P}. When comparing the results of the two
optimization methods, it is observed that both the shape of the
interface and the corner-pressure contour matches very well.
Thereby we conclude that the shape-optimization is appropriate for
the further analysis of the optimized interface.

Again it should be noted that the set of parameters used in
figures~\ref{fig:Setup_G}(b), \ref{fig:TOPOPT_P},
and~\ref{fig:Setup_fb}(b) is an extreme case, and therefore the
shape-optimization method has been further tested to ensure the
validity of this simple type of solutions.
%
\section{Comparison with alternative expansion geometries}
 \label{sec:AltGeom}
\begin{figure} 
 \centering
 \includegraphics[height=57mm]{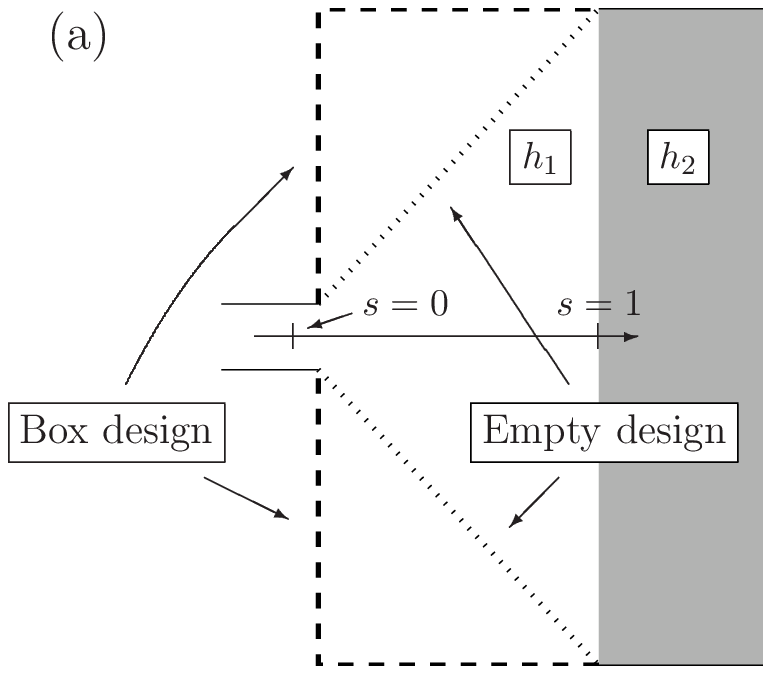}
 \hspace{1cm}
 \includegraphics[height=57mm]{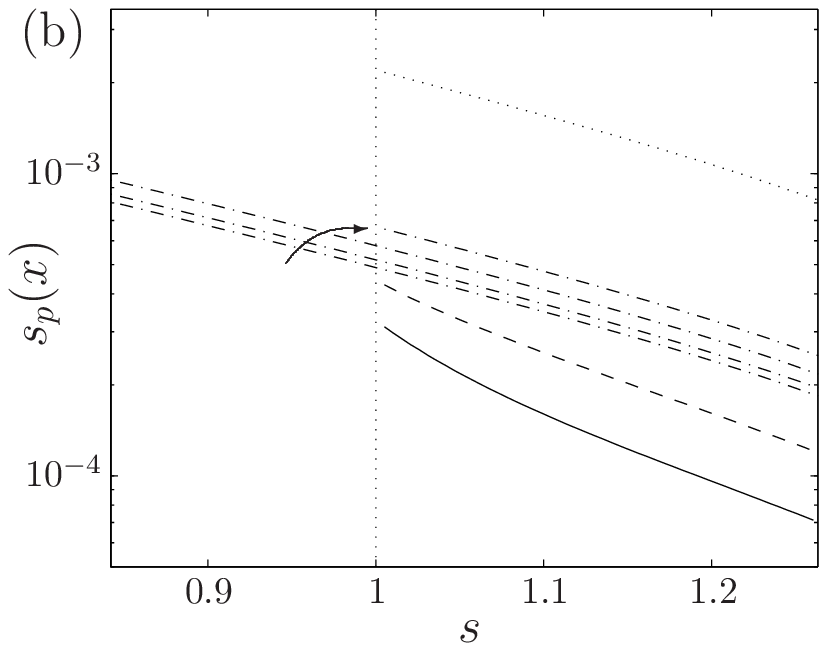}
 \caption{(a) A study of alternative expansion geometries. The
 expansion region in the \textit{Empty design} is completely filled
 with inlet height $h_1$, while the \textit{Box design} has a simple
 box of height $h_1$ as expansion region. (b) Plot of the standard
 deviation $s_p(s)$ of the pressure vertically across the expansion region
 as a function of the normalized horizontal position $s$ into the region, as seen in (a). The
 different curves are topology optimized (solid line), shape optimized (dashed line),
 empty design (dotted line), and the four types of box design (dot-dashed lines), where the arrow mark the order of the values
 $L = 0.35W, 0.55W,0.75W, 0.95W$.}
 \label{fig:AltGeom}
 \end{figure}

Now that the design of the expansion region has been optimized, it
is natural to compare its efficiency to other alternative expansion
designs. The first obvious candidate is to uniformly fill the
existing expansion region with height $h_1$ i.e.~to remove the
topology optimization distribution of height $h_2$, and this we will
call the \textit{empty design}. The next design comes as we replace
the expansion region with a simple box of width $W$ and height
$h_1$, and this will be called the \textit{box design}. Both
alternative designs are shown in figure~\ref{fig:AltGeom}(a). To
compare these new candidate designs to the optimized designs, we
will measure the homogeneity of the pressure around the end of the
expansion region. To get an quantitative measure of the homogeneity
of the pressure in the first part of the reactor, we measure the
standard deviation $\delta p(x)$ of pressure across the width along
the $y$-axis of the reactor part for a fixed $x$-coordinate:
 \beq
 \label{eq:DefSTD}
\delta p(x) = \sqrt{\left\langle\,[p(x,y)-\langle p(x,y)
\rangle_y]^2\, \right\rangle_y}
 \eeq
where $\langle\, \cdot\, \rangle_y$ is the mean along the $y$
direction. Generally $\delta p(x)$ will decrease exponentially with
the distance into the uniform reactor-part, and therefore $\delta
p(x)$ will appear as an approximately straight line when shown shown
in a log-linear plot as a function of $x$, see figure
\ref{fig:AltGeom}(b).

From the measurements presented in figure~\ref{fig:AltGeom}(b) it is
first noted that both optimized designs produce a more homogeneous
pressure-field than the alternative designs. While the empty design
give the poorest results, the box design comes closer to the shape
optimized design, and this tendency strengthen when moving the
interface closer to the inlet e.g.~for $L = 0.35W$. Since the box
design evens out the pressure due to the translation invariant
properties in the reactor part, there is a limit in how fast this
can happen, as reflected in the slope of the dash-dotted lines in
figure~\ref{fig:AltGeom}(b). On the contrary, the optimized designs
aims at homogenizing the pressure by designing the expansion-parts,
and therefore their corresponding slopes are steeper than the box
design. As a result, the optimized designs are most efficient in
quickly producing homogeneous pressure-fields.

The hydraulic resistance $R_{\mathrm{hyd}} = \Delta p/Q$ of the expansion-regions for
all the presented designs are in the range $R_{\mathrm{hyd}} = (1.7 -
3.7)\times 10^5\, \mathrm{kg}\,\mathrm{m}^{-4}\,\mathrm{s}^{-1}$,
which is five orders of magnitude smaller than the numerically
estimated $R_{\mathrm{hyd}} \approx 2.2 \times 10^{10}\,
\mathrm{kg}\,\mathrm{m}^{-4}\,\mathrm{s}^{-1}$ for the micro cell
culture \cite{Stangegaard:06a}.

The optimized designs possesses another advantage, since the
fluid-volume of the corresponding expansions regions are
significantly smaller than any of the other designs mentioned. The
fluid-volume of the different expansion-regions has been
calculated/measured and is presented in Table~\ref{tab:DesignVol}.
Also presented in the table is the volumes relative to the Shape
optimized design, and this clearly shows that extra fluid-volume is
significantly higher especially for the box designs.

\begin{table}
\caption{The volume of the different designs of the expansion chamber,
using the abbreviations: TO = Topology Optimized, SO = Shape Optimized,
ED = Empty Design, and BD-X = Box Design, with the corresponding
length-to-width fraction $X = L/W$. Second row shows the volumes in
relation to the Shape optimized design.}
\label{tab:DesignVol}
  \centering
  \begin{tabular}{|l||c|c|c|c|c|c|c|}
    \hline
    Type & TO & SO & ED & BD-0.95 & BD-0.75 & BD-0.55 & BD-0.35 \\
    \hline \hline
    Vol ($\mu$L) & 27.7 & 27.6 & 37.6 & 68.4 & 62.0 & 55.6 & 49.2 \\ \hline
    Vol/Vol(SO) & 1.01 & 1 & 1.36 & 2.48 & 2.25 & 2.01 & 1.78 \\
    \hline
  \end{tabular}
\end{table}
From the above results we conclude that the optimized designs are
generally better than the alternative designs, and we will therefore
in the following present a general description based on a vast range
of different shape optimized designs.

Knowing now the basic shape of the height interface in the expansion
region, we can now apply the right mass-conserving boundary
conditions of equation (\ref{eq:BC_HeightJump}) to the interface,
and thereby improve the model upon which the following design guide
is based.
%
\section{Design guide}
 \label{sec:Guide}
It is possible to match the numerically optimized geometry by simple
theoretical models, which only depend significantly on the
parameters $W, L, \mathcal{A}$, as the remaining parameters $\ell$
and $h_1$ only introduce minor corrections. By fitting the resulting
interface obtained in these models for given parameters to the
corresponding shaped optimized interface, we obtain an approximate
parametrization which can serve as an easily applicable guide for
practical design purposes.

 \begin{figure}[b] 
 \centering
 \includegraphics[width=14cm,clip]{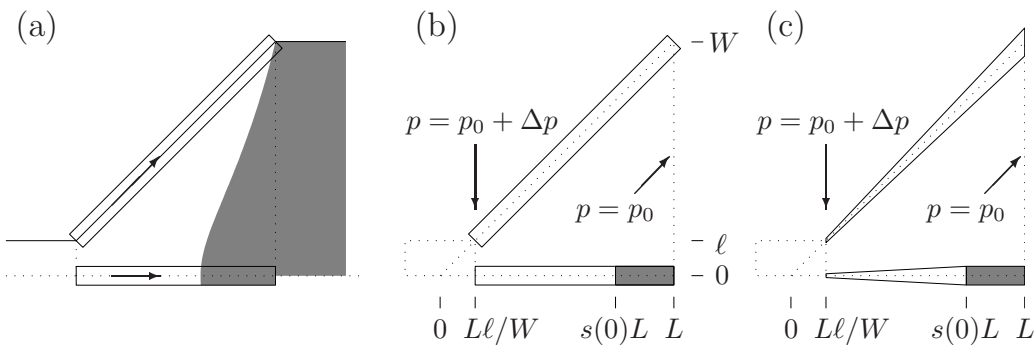}
 \caption{(a) The concept behind the models where
 the flow along two distinct flow stream in the expansion
 chamber are compared. (b) The plug flow model, and
 (c) the radial flow model. Also shown are the involved parameters
 and constraints. The gray area corresponds to regions of height $h_2$.}
 \label{fig:Model_Def}
 \end{figure}

The basic idea behind the simple models is sketched in figure
\ref{fig:Model_Def}. Given the laminar nature of the flow, we
consider an idealized narrow flow stream stretching from the inlet,
across the expansion chamber, to the entrance of the main reactor.
The first part of the stream, which is the lower part of figure
\ref{fig:Model_Def}a, goes along the horizontal symmetry-axis and
starts at the inlet where the hydraulic damping factor is $\alpha_1$
given by the height $h_1$, see equation~(\ref{eq:Lubric_rel}). Then,
at the point $(x_0,y(x_0))$ the stream hits the interface, and
continues horizontally to the point $(L,y(x_0)$ with hydraulic
damping factor $\alpha_2$. Along all streams, the hydraulic
resistance is proportional to the effective length $L_\mathrm{eff} =
\alpha_1 L_1 + \alpha_2 L_2$, where $L_1 = \sqrt{x_0^2 + y(x_0)^2}$
and $L_2 = L - x_0$ is the length of the first and second part of
the stream, respectively. Since we seek the shape $(x_0,y(x_0))$ of
the interface giving rise to the same pressure drop along the
streamlines, all streamlines must have the same effective length.
The specific form of the effective length, with its squares of $x_0$
and $y(x_0)$, then leads us to expect an expression for $y(x_0)$, or
in dimensionless form, an expression for $s(\beta)$ of the form
 \beq
 \label{eq:ParamInterf_Param}
 s_\mathrm{fit}(\beta) = 1-\frac{\mathcal{S}_{0,P}\,W^2}{L^2(\mathcal{A}\sqrt{\mathcal{A}}-1)}
     \left(1 - \frac{\sqrt{\beta^2+\mathcal{C}_P^2}}{\sqrt{1+\mathcal{C}_P^2}}
    \right).
 \eeq

\begin{figure} 
\centering
\includegraphics[width=8cm]{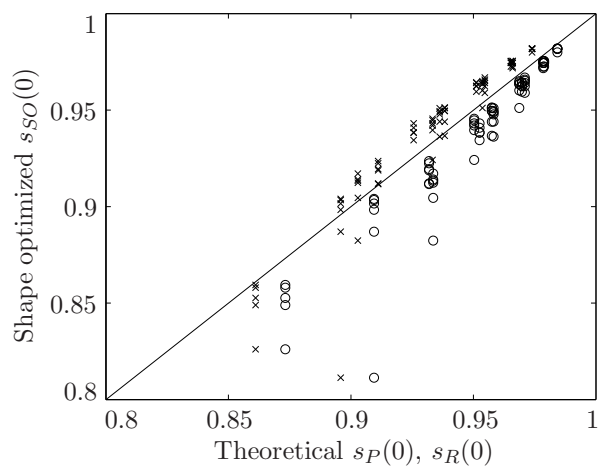}
\caption{A comparison of the calculated beginning $s(0)$ of the
transition region at the center axis between the simple model
predictions and the shape optimization model. The comparison
includes the simple model (crosses) and the radial model (circles),
and a perfect match would lie at the diagonal (solid line).}
\label{fig:Model_ShOpt}
\end{figure}

The proposed expression for $s(\beta)$ is of course not exact, but
by calculating the shape optimized interface for a large
number of parameter values, we can fit
equation~(\ref{eq:ParamInterf_Param}) and make an statistical
analysis of the obtained fitting parameters $\mathcal{S}_{0,P}$ and
$\mathcal{C}_P$. The resulting explicit parametrization becomes

 \beq
 \label{eq:ParamInterf_Value} 
 s_\mathrm{design}(\beta) = 1-\frac{2.63\,W^2}{L^2(\mathcal{A}\sqrt{\mathcal{A}}-1)}
     \left(1 - \frac{\sqrt{\beta^2+0.0729}}{1.036}
    \right).
 \eeq
This parametrization is deduced for a plug flow model, which is
shown in figure \ref{fig:Model_Def}b, while a more refined radial
model, seen in figure \ref{fig:Model_Def}c, can only be solved
numerically. A comparison between the two models, is seen in figure
\ref{fig:Model_ShOpt}. Here, the position $s(0)$ of the interface at
the center axis in the simple models for a large number of parameter
values is compared to that of the shape optimized model. These
results show no improvement by the radial model, and therefore it is
adequate to base the design guide of equation
\ref{eq:ParamInterf_Value} on the plug model.

The design guide does generally a very good job, but it maybe worth
adjusting the parameters used above ($\mathcal{S}_{0,P} = 2.63$, and
$\mathcal{C}_P = 0.27$) if the flow-homogeneity is very crucial. The
best results are obtained within the following range of parameters:
$0.4 < L/W < 1.6$,~~$6.25 <\mathcal{A}< 16$,~~$\ell/W < 0.2$, and
$h_1/W < 0.1$. This range should be met naturally for most
applications, and since we have based this work on creeping flow,
the Reynolds number of the perfusion flow should be kept below or
around unity.

In all, we thus find that equation \ref{eq:ParamInterf_Value} can
serve as a fairly accurate design guide, applicable for designing
microfluidic bio-reactors.

%
\section{Direct 3D simulation}
 \label{sec:Sim3D}
\begin{figure} 
\centering
\includegraphics[width=10cm]{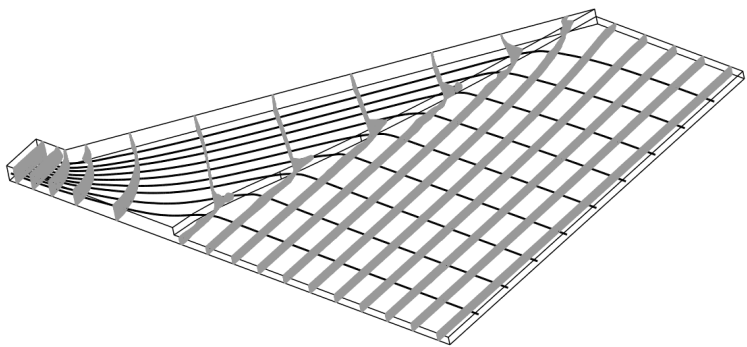}
\caption{Direct 3D numerical solution of an optimized design
following the parametrization guide of equation
\ref{eq:ParamInterf_Value}. Pressure iso-surfaces are gray, and
streamlines are solid lines. Parameters are $L = 0.95W, \ell = 0.1W,
h1 = 0.04W$, and $\mathcal{A}=4$.} \label{fig:Full3D}
\end{figure}

Up to this point we have relied on the 2D flow model based on the
Brinkman-Darcy equation. To validate this approach and test the
guideline parametrization of equation~(\ref{eq:ParamInterf_Value}),
we made a full 3D direct numerical simulations of the derived
bio-reactor design for a given set of parameters. The resulting
system was modeled and solved in \Comsol{} using a ordinary laptop
computer, and the solution is presented in figure~\ref{fig:Full3D},
where both iso-surfaces of the pressure and streamlines are showed
inside the computational domain.

Similar to the earlier quasi 3D solutions of the optimized design,
the pressure is nicely homogenized in the region of main reactor
height, and also the streamlines arrange parallel through the main
reactor. We take these results as a clear validation of the
lubrication approach used in this work. Besides, the homogeneous
flow produced by the design in figure~\ref{fig:Full3D} emphasizes
the value of derived design and the parameterizations guide of
equation~(\ref{eq:ParamInterf_Value}).

%
\section{Conclusion}
 \label{sec:Conclusion}
To increase the utilization of continuously feed microfluidic
bio-reactors, we have optimized the flow-geometry of the reactor as
to expose all immobilized organisms or substances to a very
homogeneous flow field. From this we have derived a general
guide-line of how to construct the optimal design for a broad range
of reactor-dimensions.

As the overall height of the system is much smaller than the
remaining physical dimensions, the 3D fluid flow can essentially be
described as a 2D fluid flow using a lubrication theory approach,
where an additional volume-force arise from the viscous drag by the
upper and lower channel-walls.

In this work we first achieved an optimal flow-geometry by applying
the free-form method of topology optimization. As the resulting
shape in the design had a simple single-connected topology, we subsequently applied
shape-optimization to obtain the different optimal geometries for
various reactor-dimensions. From this analysis, we have constructed
a general parametrization of an optimal design, which has been
validated by direct 3D simulations.

The design produces the homogeneous flow with a very low pressure
drop, and this will dramatically reduce the power needed to drive
the perfusion flow through the system. This opens the possibilities
of driving the perfusion in radically new ways e.g.~by buoyancy
effects. Furthermore the fluid-volume of the flow-homogenizing
design is minimized, which is essential when dealing with very
limited fluid-samples.

Besides applying the design to bio-reactors, it is also applicable
to many other microfluidics system requiring perfusion of a large
squared area, such as DNA and protein microarrays and investigation
of tissue slices using fluorescent in situ hybridization or immuno
chemistry, where samples typically are limited.

\ack F.~O.~was supported by the European Commision through the project SMART Bio-MEMS grant No. IST-016554, and by the Carlsberg Foundation, Denmark, grant No. 2006\_01\_0580.
%


\begin{thebibliography}{19}
\bibitem{Rhyd} The hydrautic resistance $R_{\mathrm{hyd}}$ of a given channel segment
is the proportionallity factor between the related pressure-drop
$\Delta p$ and the flow-rate $Q$: $\Delta p = R_{\mathrm{hyd}}\ Q$.
\bibitem{Fig_IMSO_data} The parameters used are all combinations of:
$W=0.01\,$m, $L/W = \{0.8,\,1,\,1.2,\,1.4,\,1.6\}$, $\ell/W =
\{0.05,\,0.1,\,0.2\}$, $h_1/\ell = \{0.1,\,0.4\}$, and $\mathcal{A}=
\{6.25,\,9,\,12.25,\,16\}$.
\end{thebibliography}
\end{document}